\documentclass[a4paper,nofootinbib]{revtex4}
\usepackage[pdftex]{graphicx}
\usepackage{multirow}
\usepackage{color}

\newcommand{\nc}{\newcommand}
\nc{\be}[1]{\begin{equation}\mbox{$\label{#1}$}}
\nc{\bea}[1]{\begin{eqnarray} \mbox{$\label{#1}$}}
\nc{\Section}[2]{\section{#2}\label{#1}}
\nc{\Bibitem}[1]{\bibitem{#1}}
\nc{\Label}[1]{\label{#1}}

\nc{\eea}{\end{eqnarray}}
\nc{\ee}{\end{equation}}

\nc{\bdm}{\begin{displaymath}}
\nc{\edm}{\end{displaymath}}
\nc{\dpsty}{\displaystyle}
\nc{\bc}{\begin{center}}
\nc{\ec}{\end{center}}
\nc{\ba}{\begin{array}}
\nc{\ea}{\end{array}}
\nc{\bab}{\begin{abstract}}
\nc{\eab}{\end{abstract}}
\nc{\btab}{\begin{tabular}}
\nc{\etab}{\end{tabular}}
\nc{\bit}{\begin{itemize}}
\nc{\eit}{\end{itemize}}
\nc{\ben}{\begin{enumerate}}
\nc{\een}{\end{enumerate}}
\nc{\bfig}{\begin{figure}}
\nc{\efig}{\end{figure}}

\nc{\arreq}{&\!=\!&}
\nc{\arrmi}{&\!-\!&}
\nc{\arrpl}{&\!+\!&}
\nc{\arrap}{&\!\!\!\approx\!\!\!&}
\nc{\non}{\nonumber}
\nc{\align}{\!\!\!\!\!\!\!\!&&}

\def\lsim{\; \raise0.3ex\hbox{$<$\kern-0.75em
      \raise-1.1ex\hbox{$\sim$}}\; }
\def\gsim{\; \raise0.3ex\hbox{$>$\kern-0.75em
      \raise-1.1ex\hbox{$\sim$}}\; }

\nc{\DOT}{\hspace{-0.08in}{\bf .}\hspace{0.1in}}
\nc{\Laada}{\hbox {$\sqcap$ \kern -1em $\sqcup$}}
\nc\loota{{\scriptstyle\sqcap\kern-0.55em\hbox{$\scriptstyle\sqcup$}}}
\nc\Loota{{\sqcap\kern-0.65em\hbox{$\sqcup$}}}
\nc\laada{\Loota}
\nc{\qed}{\hskip 3em \hbox{\BOX} \vskip 2ex}

\nc{\real}{{\rm I \! R}}
\nc{\Z}{{\sf Z \!\!\! Z}}
\nc{\complex}{{\rm C\!\!\! {\sf I}\,\,}}
\def\bigid{\leavevmode\hbox{\small1\kern-3.8pt\normalsize1}}
\def\id{\leavevmode\hbox{\small1\kern-3.3pt\normalsize1}}
\nc{\slask}{\!\!\!/}
\nc{\bis}{{\prime\prime}}
\nc{\pa}{\partial}
\nc{\na}{\nabla}
\nc{\ra}{\rangle}
\nc{\la}{\langle}
\nc{\goto}{\rightarrow}
\nc{\swap}{\leftrightarrow}

\nc{\EE}[1]{ \mbox{$\cdot10^{#1}$} }
\nc{\abs}[1]{\left|#1\right|}
\nc{\at}[2]{\left.#1\right|_{#2}}
\nc{\norm}[1]{\|#1\|}
\nc{\abscut}[2]{\Abs{#1}_{\scriptscriptstyle#2}}
\nc{\vek}[1]{{\rm\bf #1}}
\nc{\integral}[2]{\int\limits_{#1}^{#2}}
\nc{\inv}[1]{\frac{1}{#1}}
\nc{\dd}[2]{{{\partial #1}\over{\partial #2}}}
\nc{\ddd}[2]{{{{\partial}^2 #1}\over{\partial {#2}^2}}}
\nc{\dddd}[3]{{{{\partial}^2 #1}\over
    {\partial #2 \partial #3}}}
\nc{\dder}[2]{{{d #1}\over{d #2}}}
\nc{\ddder}[2]{{{d^2 #1}\over{d {#2}^2}}}
\nc{\dddder}[3]{{d^2 #1}\over
    {d #2 d #3}}
\nc{\dx}[1]{d\,^{#1}x}
\nc{\dy}[1]{d\,^{#1}y}
\nc{\dz}[1]{d\,^{#1}z}
\nc{\dl}[1]{\frac{d\,^{#1}l}{(2\pi)^{#1}}}
\nc{\dk}[1]{\frac{d\,^{#1}k}{(2\pi)^{#1}}}
\nc{\dq}[1]{\frac{d\,^{#1}q}{(2\pi)^{#1}}}

\nc{\bfT}{{\bf T }}

\nc{\cA}{{\cal A}}
\nc{\cB}{{\cal B}}
\nc{\cD}{{\cal D}}
\nc{\cE}{{\cal E}}
\nc{\cG}{{\cal G}}
\nc{\cH}{{\cal H}}
\nc{\cL}{{\cal L}}
\nc{\cO}{{\cal O}}
\nc{\cT}{{\cal T}}
\nc{\cN}{{\cal N}}
\nc{\cR}{{\cal R}}
\nc{\rvac}[1]{|{\cal O}#1\rangle}
\nc{\lvac}[1]{\langle{\cal O}#1|}
\nc{\rvacb}[1]{|{\cal O}_\beta #1\rangle}
\nc{\lvacb}[1]{\langle{\cal O}_\beta #1 |}
\nc{\bb}{\bar{\beta}}
\nc{\bt}{\tilde{\beta}}
\nc{\ctH}{\tilde{\cal H}}
\nc{\chH}{\hat{\cal H}}
\nc{\1}{\aa}
\nc{\2}{\"{a}}
\nc{\3}{\"{o}}
\nc{\4}{\AA}
\nc{\5}{\"{A}}
\nc{\6}{\"{O}}
\nc{\al}{\alpha}
\nc{\g}{\gamma}
\nc{\Del}{\Delta}
\nc{\e}{\textrm{e}}
\nc{\eps}{\epsilon}
\nc{\lam}{\lambda}
\nc{\Om}{\Omega}
\nc{\ve}{\varepsilon}
\nc{\mn}{{\mu\nu}}
\nc{\vp}{\varphi}

\nc{\rf}[1]{(\ref{#1})}
\nc{\nn}{\nonumber \\*}
\nc{\bfB}{\bf{B}}
\nc{\bfv}{\bf{v}}
\nc{\bfx}{\bf{x}}
\nc{\bfy}{\bf{y}}
\nc{\vx}{\vec{x}}
\nc{\vy}{\vec{y}}
\nc{\oB}{\overline{B}}
\nc{\oI}{\overline{I}}
\nc{\oR}{\overline{R}}
\nc{\rar}{\rightarrow}
\nc{\ti}{\times}
\nc{\slsh}{\hskip-5pt/}
\nc{\sm}{Standard~Model~}
\nc{\MP}{M_{\rm Pl}}
\nc{\mpl}{M_{\rm Pl}}
\nc{\tp}{t_{\rm Pl}}

\nc{\pmin}{p_{\rm min}}
\nc{\pmax}{p_{\rm max}}
\nc{\fo}{f_0}
\nc{\foi}{f_{0,i}\,}
\nc{\fop}{f_0^P}
\nc{\fou}{f_0^U}

\nc{\eff}{{\rm eff}}
\nc{\MT}{M_{\rm T}}
\nc{\ML}{M_{\rm L}}
\nc{\kk}{\vek{k}}
\nc{\pp}{{\rm p}}
\nc{\pt}{\partial_t}
\nc{\half}{{1\over 2}}
\nc{\w}{\omega}
\nc{\uhat}{\hat{U}_\w}

\nc{\etal}{\mbox{\it et al.}}
\nc{\ie}{{\it i.e. }}
\nc{\eg}{{\it e.g. }}
\nc{\trh}{T_{\rm RH}}
\nc{\ad}{{a'\over a}}
\nc{\bd}{{b'\over b}}
\nc{\Rd}{{R'\over R}}
\nc{\diag}{{\textrm{diag}}}
\nc{\mato}[1]{\tilde{#1}}
\nc{\sech}{\textrm{sech}}
\nc{\I}{\textrm{I}}
\nc{\II}{\textrm{II}}
\nc{\III}{\textrm{III}}
\nc{\vev}[1]{\langle #1 \rangle}
\nc{\hyp}{\,\; F_{1{\hskip -16pt}2}{\hskip 11pt}}
\nc{\brhom}{\overline{\rho}_M}
\nc{\brho}{\overline{\rho}}
\nc{\rhob}{\overline{\rho}}
\nc{\Pb}{\overline{P}}
\nc{\bH}{\overline{H}}
\nc{\ep}{{1+4\eps}}

\nc{\lcdm}{$\Lambda$CDM}
\nc{\lltb}{$\Lambda$LTB}
\nc{\ms}{\langle\sigma\rangle}

\def\smiley{\hbox{\large$\bigcirc$\hspace{-.80em}%
\raise.2ex\hbox{$\cdot\cdot$}\kern-.61em    
\lower.2ex\hbox{\scriptsize$\smile$}}\ }

\def\frowney{\hbox{\large$\bigcirc$\hspace{-.80em}%
\raise.2ex\hbox{$\cdot\cdot$}\kern-.635em
\lower.2ex\hbox{\scriptsize$\frown$}}\ }

\begin{document}

\title{\bf Inhomogeneity of the $\Lambda$LTB models}

\author{Peter~Sundell}
\author{Iiro~Vilja}
\affiliation{Turku Center for Quantum Physics, Department of Physics and Astronomy,
University of Turku, FIN-20014 Turku, Finland}
\date{\today}

\begin{abstract}
The Lema\'itre-Toman-Bondi (LTB) models have reported to suffer from incompatibility with cosmological observations and fine-tuning of the observer's location. Further analysis of these issues  indicates that they could be resolved by models that are compatible with the supernova Ia data, but less inhomogeneous than those that have been presented in the literature so far. We study if such models exist by employing the degrees of freedom of the LTB models in a novel manner. We discovered two scenarios which may meet the expectations, but extensive numerical and analytical investigation showed them inviable. We extended our studies to the $\Lambda$LTB models, which generalizes the LTB models by including a non-zero cosmological constant $\Lambda$ in Einsteins equations. This adds  an additional degree of freedom for the earlier scenarios and introduces a new scenario capable of meeting the expectations.  However, extensive numerical and analytical investigation reveals that inclusion of $\Lambda$ does not enhance the viability of the models.  We identify the lack of degrees of freedom to be the reason for the unviability. However, the method presented here can be generalized to models including more degrees of freedom, like the Szekeres models, which have more promise to overcome the issues in the LTB models.  
\end{abstract}

\maketitle

\section{Introduction}

The homogeneous, isotropic and spatially flat cosmological standard $\Lambda$CDM model is compatible with most of the cosmological observations \cite{Ade2013}. Examples of incompatible observations are the anomalies in the cosmic microwave background (CMB) though their significance is still under debate \cite{Schwarz2015}. Sevel authors have suggested that these anomalies may arise due to the violation of the cosmological principle. In Ref.\ \cite{AlnesAmarzquioui2006} was found that the alignment of the low multipole anomaly is not caused by the displacement of the observer from the symmetry center in a giant Lema\'itre-Tolman-Bondi (LTB) void. It is hypothesized whether the anomalous cold spot could be caused by one ellipsoidal super void or a series of consecutive smaller voids \cite{KovacsGarcia-Bellido2015}. Rotating Bianchi models have been considered as a solution to the parity-violating anomaly, but the models do not appear to contain enough rotation to do that while being compatible with the other cosmological observations \cite{Ade2014}. However, this scenario is still potential, because faster rotating models have been constructed since \cite{SundellKoivisto2015}. The list of studies executed and planned concerning the anomalies of the CMB is vastly larger than the one presented here and  is more extensively explored in Ref.\ \cite{Schwarz2015}.

In addition to the observational discrepancies, the $\Lambda$CDM model is often considered suffering from a fundamental issue: what is the nature of the cosmological constant  $\Lambda$ and why its size is what it is? The Supernova Ia (SNIa) observations made at the end of the 1990's gives the best fit for the spatially homogeneous and isotropic models with positive constant $\Lambda$ \cite{Riess1998,Perlmutter1999}, but $\Lambda$ is not a necessity for alternative models. For example, shortly after the SNIa observations, a giant LTB void was proposed to explain the SNIa observations instead of the cosmological constant  \cite{Celerier2000}. A number of studies have shown that the compatibility of different types of voids with SNIa data is comparable to that of the $\Lambda$CDM model, e.g. \cite{AlnesAmarzquiouiGron2006,Garcia-BellidoHaugbolle2008,BlomqvistMortsell2010,BiswasNotariValkenburg2010,Zumalacarregui2012,Redlich2014}.

The vast number of studies have also compiled a number of problems for the LTB models. An issue of a statistic nature is that our location in an LTB void appears to be fine-tuned close to the symmetry center of the void due to the isotropy of the CMB and the vast size of the void; more significant dipole would exist in the CMB than observed, if the observer were displaced from the symmetry center more than $\sim1$ \% of the radius of the void \cite{AlnesAmarzquioui2006,BlomqvistMortsell2010}. Issues from the observational side arise when multiple observables are fitted simultaneously. The combined data sets of SNIa and baryon acoustic oscillation (BAO) observations appears to be in conflict if the BAO data come from wide enough redshift range \cite{Zumalacarregui2012,Redlich2014,MossZibin2011} and the discrepancy increases if the BAO features of the Lyman $\alpha$ forest are included \cite{SundellMortsellVilja2015}. Moreover, tension between SNIa, local Hubble value and CMB have been reported \cite{BiswasNotariValkenburg2010,Redlich2014,MossZibin2011} and the authors in Ref.\ \cite{Garcia-BellidoHaugbolle2008b} found that kinematic Sunyaev-Zel’dovich  (kSZ) effect give stringent constraints for the size of the void so that the widest and the most underdense voids are ruled out. On the other hand, the aforementioned studies have suppressed the degree of freedom of the inhomogeneity of the bang time function by assuming it to be a constant; the benefit of this is that the well-known homogeneous early time evolution can now be employed for the models. Inhomogeneous bang time is a significant degree of freedom, e.g. it has been shown that the SNIa data can be modelled without a void but using inhomogeneous bang time instead \cite{Krasinski2014} and that the aforementioned incompatibility between SNIa, local Hubble value and CMB can be removed by inhomogeneous bang time \cite{Clifton2009,BullCliftonFerreira2012}. Nevertheless, the simultaneous fitting of the SNIa, the small-angle CMB, the local Hubble rate, and the kSZ effect appears to rule out the models due to the amount of inhomogeneity \cite{BullCliftonFerreira2012}. We would like to draw the attention into two aspects of the above studies. The SNIa alone and the combined SNIa, local Hubble value and CMB data appears to favor inhomogeneity while other evidence points to homogeneity or moderate inhomogeneity. In addition, all the aforementioned studies employ the degrees of freedom using  ansatzes of similar type, hence the results are not general and different results may appear with different ansatzes. Particularly, ansatzes which aim to minimize the inhomogeneity so that the model would still remain compatible to the SNIa data have not been considered. We aim to fill this gap. 

We search for less inhomogeneous LTB models compared with those usually presented in the literature by employing the idea presented in \cite{SundellVilja2014}.
Consider an observable emitting light at $r_e$ and we measure its redshift to be $z_e$. If we measure the redshift of the same observable at some later time and obtain the same value $z_e$, the redshift does not change in observer's time, i.e. the redshift drift is zero. The SNIa surveys measures the luminosity distance, $D_L$, with respect to redshift. Therefore, if the redshift does not change in observer's time neither does the luminosity distance. If the redshift drift remains zero, we do not know for how long the observable at $r_e$ has emitted light that has reached us with redshift $z_e$. Hence, we do not know  the age of the universe at $r_e$, albeit we can give the minimum age for a given model and initial conditions.  In the LTB models, the time of birth of the universe is described by the bang time function $t_b(r)$, which is dependent on the coordinate distance, thus we can merely evaluate $t_b^{min}(r_e)$ corresponding the minimum age of the universe at $r_e$ for a given model and initial conditions. If this takes place for all observables at some range of redshift, we can evaluate the shape of  $t_b^{min}(r)$  which corresponds the minimum age of the universe for all $r$ at that range for a given model and initial conditions.  Consequently, we have the freedom to choose the function $t_b(r)$ as long as it does not make the universe younger than  $t_b^{min}(r)$ at any $r$ in the given range. If this takes place for an LTB model that is close to homogeneous except by the part of the bang time function (like in Ref. \cite{Krasinski2014}), the luminosity distance of the model is not affected by changing the bang time function  as long as it makes the universe older. This way we may obtain e.g.\ homogeneous bang time.

On the other hand, if one ignores the conceptual issues of $\Lambda$, it is straightforward to introduce the cosmological constant into the LTB models, these models are called the $\Lambda$LTB models. These models include the $\Lambda$CDM as a special case and therefore can be used to study the effects of  radial inhomogeneity with respect to the $\Lambda$CDM model more extensively than using  the perturbation theory of the spatially homogeneous and isotropic  space-times. In addition, the $\Lambda$LTB models allow us to explore the nearly LTB cases where the cosmological constant is nearly zero. This possibility was utilized in Ref. \cite{MarraPaakkonen2010}, where was evaluated which $\Lambda$ value is preferred in the context of the $\Lambda$LTB models when fitting SNIa, CMB, local Hubble, and BAO data; the $\Lambda$ corresponding to the $\Lambda$CDM model was favored though their analysis could be improved in number of ways (which they list in their conclusions). The $\Lambda$LTB model  have been used to test the  Copernican principle, but the observations are not accurate enough to confirm the principle \cite{Valkenburg2014}. More conservative  approach was not able to rule out  $\sim 15\%$ inhomogeneity in the  matter density \cite{Redlich2014}.
Nevertheless, none of these studies have employed the degrees of freedom as we shall do in this paper, hence explore these scenarios also by expanding our investigations into   the $\Lambda$LTB models.

The novel manner introduced in this paper to employ degrees of freedom in the $\Lambda$LTB models can be applied into more general inhomogeneous models, like  the Szekeres models. Furthermore, it appears that the issues in the LTB models do not simply vanish by considering more general models. It have been shown that the latter class of models allows the observer to be displaced further away from the center of the void than the LTB models \cite{BolejkoSussman2011}. However, this does not necessarily solve the fine-tuning issue because (at least in the quasispherical models) the locus of the observer appears still to be fixed for each Szekeres model, it is just not fixed to be in the center of the void \cite{BuckleySchlegel2013}.  Hence, there is also a need for applying the method to the Szekeres models. It can also be a cumbersome task to give ansatzes for all the free functions in the Szekeres models and then fit the data. As our approach yields constraining equations intrinsically, it can be used as a theoretical tool to probe more general models before data fitting procedures. The  development toward more realistic and exact models of the universe proceeds continuously. For example, in Ref.\ \cite{SussmanGaspar2015} the authors have shown that the universe can be described by a mesh of “pancake” shaped overdensities and voids whose evolution is physical and can be traced back to the early universe where the initial conditions match the standard conditions in the early universe. In Ref. \cite{SussmanGasparHidalgo2015} this is result used to generalize the model in \cite{BolejkoSussman2011}, where the “pancakes” are $\sim 100$ Mpc configurations today and are distributed around a central spheroidal void.

The paper is organized as follows. In Section \ref{LTBmodels} we introduce the aspects of the $\Lambda$LTB models relevant to this work. In the following Section \ref{mechanism} we introduce the method we shall employ  to reduce the inhomogeneity of the $\Lambda$LTB  models which are compatible with SNIa data. The specific models obtained by employing the method are presented and analyzed in Section \ref{models}. The results are concluded in Section \ref{conclusions}.

We use units in which the speed of light $c=1$ and km s${}^{-1}$ Mpc${}^{-1}=1$.

\section{The Lema\^{i}tre-Tolman-Bondi models} \label{LTBmodels}

The LTB models \cite{Lemaitre,Tolman,Bondi} describe an inhomogeneous 
but isotropic universe that obeys the Einstein equations and is sourced by dust. Hence, the energy momentum tensor reads as
$T^{\mu \nu}=\rho u^{\mu} u^{\nu}$, where $\rho=\rho(t,r)$ 
is matter density and $u^{\mu}$ is the local four-velocity and $u^{\mu}=\delta^{\mu}_t$  as the coordinates are 
assumed to be comoving. We consider the special case of the model where the observer is at the origin.\footnote{We aim to address to the issues by constructing solutions which are less inhomogeneous than those previously presented in the literature. If such scenarios cannot be found within the case where the observer is at the origin, it is unexpected to find them from the off-center solutions either.}
The metric in the LT model in the  standard synchronous gauge  is written as
\be {LT_metric}
ds^2=-dt^2+\frac{R_{,r}^2 dr^2}{1+2 e(r) r^2}+R^2 \left( d \theta^2 +
\sin^2 \theta\, d\phi^2 \right),
\ee
where $R=R(t,r)$, $2e(r)r^2\geq -1$ and the subscript after comma denotes a partial derivative, $R_r=\partial R(t,r)/\partial r$. The 
evolution of the universe is described by the local scale factor $R$, whereas function $e$
represents the local curvature. The cosmological constant $\Lambda$ is straightforward to include
in the Einstein equations, which yield the differential equations for the $\Lambda$LTB models:
\be {R1}
R_{,t}^2=\frac{2M(r)}{R}+2 e(r) r^2+\frac{\Lambda}{3}R^2,
\ee 
and 
\be {rho}
\kappa \rho=\frac{2M_r(r)}{R_r\,R^2},
\ee
where $R_{,t}=\partial R(t,r)/\partial t$, $M(r)$ is an arbitrary function, $\kappa=8\pi G$, and $G$ is the 
Newton's constant of gravity. 

In the absence of the cosmological constant, i.e.\ in the case of the LTB models,  the solution of Eq. (\ref{R1})  for positive $e(r)$ is
\be{R}
R=\frac{M(r)}{2e(r) r^2}\left( \cosh[\eta]-1\right), \qquad \sinh[\eta]-\eta=\frac{[2e(r)]^{3/2}r^3}{M(r)}[t-t_b(r)],
\ee
where $t_b(r)$ is the bang time function. For $e(r)=0$ the solution is
\be{R0}
R=\left\{\frac{9}{2}M(r)[t-t_b(r)]^2\right\}^{1/3},
\ee and
for negative $e(r)$ it is
\be{R-}
R=-\frac{M(r)}{2e(r) r^2}\left( 1-\cos[\eta]\right), \qquad \eta-\sin[\eta]=\frac{ [-2e(r)]^{3/2}r^3}{M(r)}[t-t_b(r)].
\ee

The redshift along the null geodesic is given as \cite{BKHC2010} 
\be  {rs}
\frac{dz}{dr}=\frac{(1+z)R_{,rt}|_{ng}}{\sqrt{1+2e(r)}},
\ee 
where $|_{ng }$ denotes that the function is evaluated along the null geodesic, where $ds^2=d\theta^2=d\psi^2=0$. That is,  $R_{,rt}|_{ng}=R_{,rt}(t(r),r)$, where function $t$ is solved from the differential equation for incoming light rays:
\be {ng}
\frac{dt}{dr}=- \frac{R_{,r} }{\sqrt{1+2e(r)}}.
\ee

The redshift drift describes how the redshift $z_e$ of an observable at $r_e$ changes in observers time $t_0(=t(0))$ and is given as \cite{SundellVilja2014}
\bea{rsd}
\frac{dz_e}{dt_0}=(1+z_e)\int_{ 0}^{r_e}\frac{R_{,ttr}|_{ng}}{\sqrt{1+2e(r)} }\frac{1}{1+z}  dr   ,
\eea
where 
\be{Rttr}
R_{,ttr}=-\frac{M_{,r}(r)}{R^2}+2 \frac{M(r)}{R^3}R_{,r}+\frac{\Lambda}{3} R_{,r}.
\ee
Redshift is necessarily evaluated along the null geodesic and hence we have suppressed the notation $|_{ng}$  in conjunction with $z$.

The apparent horizon (AH) is the hypersurface in space-time where $dR(t|_{ng},r)/dr=0$ \cite{BKHC2010}.
Taking the derivative and substituting Eq.\ (\ref{ng}) in yields the  AH condition
\be{AHc}
2M(r)+\frac{\Lambda}{3}R|_{ng}^3-R|_{ng}=0.
\ee

In this paper, we shall use the gauge
\be{gauge}
M(r)=M_0r^3.
\ee
 However, instead of choosing a value for the constant $M_0$, we fix  $R(t_0,r)/r|_{r=0}=1$ to fully employ the gauge degree of freedom. The benefit of this  can be seen by consulting Appendix \ref{originconditions} and rewriting Eq.\ (\ref{R1}) at the origin as
\be{R10}
H_0^2=2M_0+2e_0+\frac{\Lambda}{3},
\ee
where $e_0=e(0)$, $H_0=H(t=t_0,r=0)$ and $H(t,r)=R_{,t}(t,r)/R(t,r)$. This equation has evident resemblance to the Friedmann equation and therefore gives intuitive conception for the constants $M_0$, $e_0$ and $\Lambda$.

Even though the more modern parametrizations, the q-scalars and their fluctuations, definitely have their benefits \cite{Sussman2013,Sussman2015}, we shall use  the old metric variables (Eq. (\ref{LT_metric})) for the following reason. This paper aims to find a method of reducing the amount of inhomogeneity in the LTB models (also by including $\Lambda$). The studies motivating this paper (presented in the Introduction) have  used different metric variables and gauges, which makes it cumbersome to define the amount of the inhomogeneity in these models. In this paper, the approach to confront the SNIa data similar to that in Refs. \cite{SundellMortsellVilja2015,Krasinski2014,Krasinski2014b} and because  the origin conditions are simpler to process using the same variables as in the two latter ones,  we choose to use the same metric variables as they do. 

\section{the method of smoothening the  inhomogeneity in the $\Lambda$LTB models}\label{mechanism}

For the reasons described in Introduction, we aim to reduce the inhomogeneity of the LTB models compared with those previously presented in the literature while still remaining compatible with SNIa observations. Below we shall describe the method we use to obtain such smoothed inhomogeneity and give a quantitative measure for the allowed size of  inhomogeneity. We shall also discuss the observational capabilities in the future observations, which we shall use later when presenting our results.

Consider an observable at $r_e$ emitting light which we observe redshifted as $z_e$. If the system is approaching  a state where the redshift sets in a constant value at $r_e$ and $z_e$ is close to that value, we may not be able to distinguish the evolving of the redshift due to our limitations in cosmological observations. If this takes place for all observables at some range of redshift (or coordinate distance), the observed luminosity distance, $D_L(z)$, covering that range would also remain unaltered as time passes. Consequently, we cannot predict the form of the bang time function covering the range in question due to the uncertainties in the observations and therefore this scenario adds freedom to adjust the bang time function according to other observations. For example, it has been shown that the SNIa data can be modeled using an LTB model where the curvature is homogeneous but the bang time is not \cite{Krasinski2014}. If the redshift drift would be approximately zero at the redshift range covering the SNIa sample, changing the bang time function would not alter the luminosity distance enough to be observationally significant.

 In the LTB space-times, the angular diameter distance, $D_A$, is related to the luminosity distance as $D_L(z)=(1+z)^2 D_A(z)$ \cite{Etherington1933} and $D_A(z)=R(t(z),r(z))=R|_{ng}$, where $t(z)$ and $r(z)$ are solved from Eqs.\ (\ref{rs}) and (\ref{ng}). 
To ensure the fit of our model  to SN Ia data, we impose the LTB model to have the Friedmanian angular diameter distance
\be{DA}
D_A=\frac{1}{(1+z)H_0^F} \int_{0}^{z}\frac{dz'}{\Omega_m^F(1+z')+\Omega_{\Lambda}},
\ee
which for $\Omega_m^F=0.295$ and $\Omega_{\Lambda}^F=1-\Omega_m^F$ is the  luminosity distance of the best fit flat $\Lambda$CDM model for SNIa data \cite{Betoule2014}. Different studies give different best fit values for the  Friedmannian Hubble  $H_0^F$. For example, the local Hubble measurement in Ref.\ \cite{Riess2011} gives the best fit $H_0^F=73.8 \pm 2.4$ km s${}^{-1}$ Mpc${}^{-1}$ whereas the Planck collaboration \cite{Ade2015} reports a lower best fit value $H_0^F=67.8 \pm 0.9$ km s${}^{-1}$ Mpc${}^{-1}$. In this paper, we shall use the value  $H_0^F=70$ km s${}^{-1}$ Mpc${}^{-1}$ unless otherwise noticed.\footnote{The exact value of $H_0^F$ turns out not affecting the results qualitatively as our numerical investigations indicate that the results are not sensitive to $H_0^F$ value.} 

Imposing $R$ to be the angular diameter distance (\ref{DA}) and taking its  derivative along the null geodesic with respect to $r$, we find
\be{DAr}
\frac{d R|_{ng}}{dr}=\frac{d D_A}{dz}\frac{d z}{dr}.
\ee
On the other hand, along the null geodesic we also have
\be{dRdr}
\frac{d R|_{ng}}{dr}=R_{,t}|_{ng}t_{,r}|_{ng}+R_{,r}|_{ng}.
\ee
When $\Lambda=0$, the differential equations (\ref{rs}), (\ref{ng}), (\ref{DAr}) and (\ref{dRdr}) consist of the functions $M(r), \,e(r),\,z(r),\, R(r),\, t(r)$, $t_b(r)$ and their derivatives with respect to $r$ along the null geodesic. Employing Eq.\ (\ref{gauge}) to fix the gauge degree of freedom, we still need one more equation to close the system. We shall choose this equation according to the situation. When $\Lambda \neq 0$, we cannot give $R(t,r)$ in algebraic form and as a consequence the equations are no longer dependent on $t_b$, but instead on $R_{,r}$. 
We use Eq.\ (\ref{dRdr}) to eliminate $R_{,r}$ from the equations. Nevertheless, an extra equation is required to close the system also in this situation and we shall choose this equation accordingly.

\subsection{Vanishing redshift drift}

The redshift drift can vanish only if the integral in Eq.\ (\ref{rsd}) is zero. This is true for all positive $r\leq r_e$ only if $R_{,ttr}|_{ng}=0$ for all $0\leq r\leq r_e$. However, in Appendix \ref{originconditions} we show that $R_{,ttr}|_{ng}$ is zero at the symmetry center only for special conditions and in general $R_{,ttr}|_{ng} \neq 0$ there. On the other hand, the SNIa sample used in Ref.\ \cite{Betoule2014} covers the redshift range $0.01 < z < 1.2$, hence the redshift drift does not necessarily need to be zero for $z<0.01$. Consequently, it is sufficient to require that both $R_{,ttr}|_{ng}$ and the integral in the redshift drift equation become zero at some critical value $r_c$, corresponding to the critical redshift $z(r_c)\equiv z_c\leq0.01$, and $R_{,ttr}|_{ng}$ remains zero for $r>r_c$. This ensures that also the integral is zero for $r>r_c$. 
Therefore, it is convenient to divide the redshift drift Eq.\ (\ref{rsd}) into two parts as
\be{rsd2}
\frac{dz}{dt_0}=(1+z)\left[ F_{0c}+F_{ce} \right],
\ee 
where we have defined
\bea{F0c}
F_{0c}&\equiv & \int_0^{r_c}\frac{R_{,ttr}|_{ng}}{\sqrt{1+2e(r)r^2}\,(1+z)}dr,  \qquad  0\leq r_c ,\,\, R_{,ttr}(t|_{ng},r)|_{r=r_c}=0, \,\, F_{0c}(r_c)=0, \\ \label{Fce} F_{ce}&\equiv&\int_{r_c}^{r_e}\frac{R_{,ttr}|_{ng}}{\sqrt{1+2e(r)r^2}\,(1+z)}dr,  \qquad  r_c \leq r \leq r_e ,\,\, R_{,ttr}|_{ng}=0.
\eea
This division enables us to investigate the qualitatively different parts, $F_{0c}$ and $F_{ce}$, separately.  

 If $F_{ce}$  cover the same range of redshift as the SNIa sample,  the redshift drift can vanish there too.  We employ the results of  Ref.\ \cite{Betoule2014}, where the range is $0.01<z<1.2$. The function of the  term $F_{0c}$ is to ensure zero redshift drift  at some $r_c$ and is unnecessary if $r_c=0$. In Appendix \ref{originconditions} we show that this takes place when 
\be{Rttron0}
M_0=\frac{\Lambda}{3}.
\ee

Strictly speaking, it is not a necessity that $R_{,ttr}|_{ng}$ is exactly zero at  $0.01<z<1.2$, but it is sufficient that $R_{,ttr}|_{ng}$ tends to zero and is close enough along the current null geodesic due to our limitations in observational accuracy. By close enough we here mean that our observational limitations would no longer be able to distinguish the evolution of  $D_L(z)$. 

\subsection{The amount of inhomogeneity}

We aim to find less inhomogeneous LTB models compared with those usually presented in the literature. The amount of inhomogeneity is cumbersome to define because different studies have used gauges and different functions to characterize the inhomogeneity. Nevertheless, it appears that at least one of the functions characterizing the radial inhomogeneity has tens of per cents difference between its minimum and maximum values in the models presented in the literature. For example, in Ref. \cite{AlnesAmarzquioui2006}, where the fine-tuning issue was first acknowledged, two models are considered and the inhomogeneity of their models can be characterized by dividing the relative matter density outside the underdensity by its value at the center, which for their Model I gives 5 and for their Model II gives 4. In Ref. \cite{BlomqvistMortsell2010}, where the fine-tuning issue was confirmed for a different type of void, two different sets of best fit parameters were found corresponding to two different SNIa samples. Dividing the relative matter density outside the underdensity by its value at the center  in this case yields the values 4.6 and 7.7 for the different SNIa data sets. The contrast in both studies is hundreds per cents, which appear typical for the papers where the LTB model is used to model the universe from the surface of last scattering to the present time. 
Naturally, if a narrower redshift range covered, the inhomogeneity contrast reduces. For example, the authors in Ref. \cite{BlomqvistMortsell2010} found the best fit values of the inside and the outside matter densities of the void for one of the SNIa samples to be 0.16 and 0.29, respectively, hence their ratio is 1.81 and the inhomogeneity is only tens of per cents. However, large inhomogeneity is also used to model SNIa data: in Ref.\ \cite{SundellMortsellVilja2015} the ratio of the present matter density corresponding to redshifts 1.6 and 0 is 4.7 whereas the author in Ref.\ \cite{Krasinski2014b} found $e(z=1.6)/e(0)\approx 0.18$ for the same redshifts. We note that these two results corresponds to the same physical system, the only difference is the chose gauge and the functions characterizing the inhomogeneity. Consequently, these numbers, or actually $1/0.18\approx 5.6$ and 4.7, is a representation of the effect of the gauge choice and whether we compare matter density or curvature. 
In the case of non-zero cosmological constant, the conservative approach used in Ref.\ \cite{Redlich2014} was not able to rule the inhomogeneity of $\sim 15\%$ out with observations, but less conservative approach was able to lay more stringent constraints \cite{Valkenburg2014}.

In this paper, we concentrate our investigations on the range  $z \in [0,1.2]$ and based on the above discussion we restrict our studies by constraining the curvature as 
\be{econtrast}
\abs{\frac{e|_{z=1.2}-e_0}{e_0}}\lsim0.1, 
\ee 
independently on the value of the cosmological constant. This constraint appears quite stringent, because the models used here have the same luminosity distance - redshift relation as in the models used in Refs.\ \cite{SundellMortsellVilja2015} and \cite{Krasinski2014b}, where $e|_{z=1.6}/e_0=0.18$. Therefore, we allow a moderate violation of the constraint.

\subsection{Observational prospects}

The redshift drift predicted by the $\Lambda$CDM model should be in our observational reach within decades. The Lyman-$\alpha$ forest appear to be the best candidate \cite{Loeb1998}, \cite{Liske2008} for the accurate measurement of the frequencies of visible light. These observations cover the high redshift $z\gsim2$ region, whereas the radio sources for 21 cm hydrogen absorption systems can be used to obtain the redshift drift for $z<2$ \cite{Yu2014}. The expected resolution in these future surveys is of the order
\be{resolution}
\frac { dv}{dt_0} \sim 10^{-1}\, \frac{\text{cm}}{ \text{s} \times \text{yr}},
\ee
where the acceleration of a given object is  defined as
\be{svs}
\frac { dv}{dt_0} \equiv\frac{c }{1+z_e}\frac{dz}{dt_0} .
\ee
We shall present our results in the units of cm/s/yr to make them comparable with our observational capabilities.

\section{The models}\label{models}

The redshift drift  converges to zero, if  $F_{ce}\rightarrow0$ as $t\rightarrow \infty$. Furthermore,  $F_{ce}$ tends to zero when $R_{,ttr}|_{ng}$ does, which can occur in three scenarios that we refer to as the big universe, the static universe, and the bizarre universe.
  By the big universe we mean the scenario where 
\be{bu}
R|_{ng} \rightarrow \infty, \quad \Lambda=0,
\ee 
as $t\to \infty$ for all $r\in [r_c,r_e]$. The big universe is of interest because it  represents the LTB cases. 
The static universe is the case where $R|_{ng}$ and $R_{,r}|_{ng}$ approach to functions $c_1=c_1(r)$ and $c_2=c_2(r)$ as
\be{su}
R|_{ng} \rightarrow c_1, \quad R_{,r}|_{ng}\rightarrow c_2, \quad -\frac{M_{,r}(r)}{c_1^2}+2 \frac{M(r)}{c_1^3}c_2+\frac{\Lambda}{3}c_2=0,
\ee
as $t\to \infty$ for all $r\in [r_c,r_e]$. This scenario includes the $\Lambda$CDM model as a special case. In Ref.\ \cite{SundellVilja2014} was shown that the $\Lambda$CDM model contains no location $r_e$ where the redshift drift remains zero, hence it is interesting to find out if this would change for by introducing small inhomogeneity using the $\Lambda$LTB models. Furthermore, this case includes also pure LTB models (when $\Lambda=0$), but these scenarios differ from the ones in the big universe as here the negligible redshift drift is obtained by a suitable cancellation in $R_{,ttr}|_{ng}$, whereas in the big universe all the components decreases individually.
The bizarre universe is characterized by the conditions
\be{biu}
R|_{ng} \rightarrow \infty, \quad R_{,r}|_{ng}\rightarrow 0, \quad \frac{R_{,r}|_{ng}}{R|_{ng}^3}\rightarrow 0
\ee
as $t\to 0$ for all $r\in [r_c,r_e]$. This universe is bizarre because it is expanding in the angular direction while contracting in the radial direction.

\subsection{The static universe}

As defined in Eq.\ (\ref{su}), by the static universe we refer to the case where  $R$ and $R_{,r}$ tend to constant values for all $r\in [r_c,r_e]$ so that the terms on the right hand side of Eq.\ (\ref{Rttr}) suitably cancel at the asymptote. From  Eq.\ (\ref{R1}) we see that $R$ can approach zero while time tends to infinity. Consider an observable emitting light at $r_e$. The angular diameter distance to that object, $D_A(r_e)$,  is obtained from Eq.\ (\ref{R1}) as
\be{T}
\int_{t_b(r_e)}^{t(r_e)}dt=\int_0^{D_A(r_e)}\frac{dR}{\sqrt{2M(r)/R+2e(r)r^2+\Lambda R^2/3}}.
\ee
 As the integral on the  left hand side grows to infinity,  the integral on the right hand side tends to a constant if  the square root converges to  zero. For a growing $R$, this can take place only if $e$ or $\Lambda$ or both are negative.

As mentioned above, one more equation is required to close the system. We shall impose the equation
\be{f}
\frac{R_{,ttr}|_{ng}}{\sqrt{1+2e(r)} }\frac{1}{1+z}=f(r),
\ee
 where $f(r)$ is a step function
\be{fdef}
f(r) =
  \left\{
  \begin{array}{ll}
 f_0 -\frac{f_0}{\beta}r&  r \leq \frac{3}{2}\beta\\
- f_0 + \frac{f_0}{3\beta}r \quad & \frac{3}{2}\beta<r\leq 3\beta \\
0 &  3\beta<r,
  \end{array}
\right.
\ee 
where we choose the parameter $\beta$ so that $z_c=0.01$ and $f_0=R_{,ttr}(t_0,0)$.

The set of Eqs.  (\ref{rs}), (\ref{ng}), (\ref{gauge}), (\ref{DAr}),   (\ref{dRdr}) and  (\ref{f}) contains the undetermined free parameters\footnote{As shown in Appendix \ref{originconditions}, the value of $e_0$ is  determined if $M_0$, $\Lambda$, and $H_0^F$ are given, hence we do not consider $e_0$ here as a free parameter.}   $\Omega_m^F$, $H_0^F$ $\beta$, $M_0$ and $\Lambda$. We numerically solve the system by first fixing $\Omega_m^F$, $H_0^F$ and $M_0$ (or $\Lambda$) and then using trial and error to find the correct values for $\beta$ and $\Lambda$ (or $M_0$) so that $z_c=0.01$ and the AH condition (\ref{AHc}) is satisfied. We found that for each set $(\Omega_m^F, H_0^F,M_0$) (or  $(\Omega_m^F, H_0^F,\Lambda$)) there is only one set ($\beta,\Lambda$) (or $(\beta,M_0)$) that met the conditions. 

We imposed $\Omega_m^F=0.3$ and $H_0^F=70$ and  studied the system numerically by varying the free parameters   $\Lambda$, $M_0$ and $\beta$; the results  are given in Table \ref{LTBt0}. The cosmological constant is constrained from above for positive $M_0$: $\Lambda \lsim 1.1 (H_0^F)^2$. The maximum value  $\Lambda \sim 1.1 (H_0^F)^2$ corresponds to  small $M_0$ values and $\Lambda$ decreases as $M_0$ increases.  For  $\Lambda \sim 1.1 (H_0^F)^2$, the AH condition is considerably more sensitive to the value of $\Lambda$ compared with the value of $M_0$; only fractions of change in $\Lambda$ imposes orders of magnitude changes in $M_0$, as can be seen on Table \ref{LTBt0}.  This sensitivity difference level off when $M_0$ increases and turns around when $M_0$ is of the order $\sim10^4 (H_0^F)^2$. However, the inhomogeneity of  $e(r)$ increases along with   $M_0$, thus  we study only the case of non-negative cosmological constant further.

\begin{table} \caption{The results of the numerical analysis of the static universe model (Eq.\ (\ref{su})) characterized by  $\Omega_m^F=0.3$ and $H_0^F=70$. The AH is located at the angular diameter distance $R_{AH}=1.74472$ Gpc and the AH condition is satisfied for each set of parameters by the accuracy $\sim 10^{-6}$. The values of $\Lambda \chi^2 +2e$  corresponds to coordinate distances at $ r_c \leq r\leq r_{AH}$ and $3\beta=r_c$.} \label{LTBt0} 
\begin{center}
  \begin{tabular}{| c | c | c | c | c | c | c | c |} 
    \hline 
 $\Lambda/(H_0^F)^2$  &   $M_0/(H_0^F)^2$  & $e_0/(H_0^F)^2$ &  $e_{AH}/(H_0^F)^2$ & $r_{AH}\quad$ [Gpc] &  $\Lambda \chi^2+2e(r)$  & $z(r_c)$ & $\beta \quad$ [Mpc]\\ \hline
 $1.100446$ &  $3.123000 \times 10^{-4}$  & 0.31628 &0.00215794  &36.3663&   $\geq 44.2067$ & $0.0100412$ &  $150$\\ \hline
 $1.10012(=3M_0)$ & $0.366706$  & $-0.050059$  & $0.240929$  & $3.44587$    &  $\geq 4900$ & 0 & -\\ \hline
 $1.1001017$ & $0.5$  & $-0.18335$  & 0.295413  & 3.10873&   $\geq 6053$ &0.0100387 & 12.8  \\ \hline
$0.82$ & $75863.4$  & $-75863.0$  &$1105.78$ &0.0586031 &   $\geq 311534.$ &0.0100121 & $0.459$  \\ \hline
 $0.81$ & $75802$  & $-75801.6$  &$1120.25$ &0.0585791 &   $\geq -1607.35$ &0.0100486 & $0.465$  \\ \hline
$0.001$  & $89756.38$ & $-89755.9$  & $2112.78$  & $0.0562716$  &  $\geq-3.10893\times 10^{7}$  & $0.0100328$ & $0.66$\\    \hline
  $0$ & $89798.1$  & $-89797.6$ & $ 2118.09$ & $0.0562408$  &  - & $0.0100309$ & $0.66$   \\ \hline
 $-0.001$  & $89839.92$  & $-89839.4$  & $2115.72$  &  $0.0562600$  & - & $0.0100289$ &  $0.66$   \\ \hline
$-3$  & $94741.78$  & $-94740.8$  & $6279.34$  &  $0.0581723$ &  - & $0.010023$ &  $0.965$\\    \hline
  \end{tabular} 
\end{center} 
\end{table}

For $\Lambda >0$ we need $e<0$ to enable $R$ to converge to a constant  for all $r\in [r_c,r_e]$.
It turns out convenient to define a  parameter $\chi$ as
\be{chi}
\chi^3=3\frac{M_0}{\Lambda},
\ee
because it can be used to characterize the value of $R_{,ttr}$ at the origin (compare Eqs.\ (\ref{Rttron0}) and (\ref{chi})) and it can be used to evaluate whether a model can converge on a state where $R$ tends to a constant value. The latter is obtained as follows. Consider an observable at a constant radial coordinate distance $r_e$. As $R$ increases from zero in time, $R_{,t}^2$ (which is the function inside the square root in Eq.\ (\ref{T})) takes its minimum value at
$$
R=\chi r.
$$
Substituting this and (\ref{chi}) into  $R_{,t}^2$ yields its minimum value $R_{,t}^2|_{min}= \Lambda \chi^2+2e$. If $ \Lambda \chi^2+2e$ is positive, $R$ can not converge to a constant, hence the condition for $R$ to approach a constant value is
\be{ineq2}
 \Lambda \chi^2+2e \leq 0.
\ee

Table \ref{LTBt0} manifests that this condition is satisfied only in the cases where the condition for maximum inhomogeneity (\ref{econtrast}) is severely violated. The lowest $\Lambda$ for which the condition (\ref{ineq2}) is satisfied for some $r \in [r_c,r_{AH}]$ lies at $0.81(H_0^F)^2<\Lambda<0.82(H_0^F)^2$. The exact value is dependent on $\beta$ and how accurately the condition $z_c=0.01$ is met using the numerical methods. These results are unexpected for two reasons. First, we should not obtain any solutions by demanding that the redshift drift is exactly zero for $z\in [0.01,1.2]$, because, according to our reasoning, this can take place only if the right hand side of Eq.\ (\ref{T}) is singular. Second, we presupposed that $dz/dt_0 \to 0$ if $R$ tend to constant, but the cases where $\Lambda> 0.81 (H_0^F)^2$ represents counter examples. We reanalyzed the scenario by considering approximately zero redshift drift for $z\in [0.01,1.2]$, but the results did not considerably differ from the exact zero case. This can be understood by plotting $R_{,t}|_{ng}$ (which is the square root in Eq. (\ref{T})): one would expect $R_{,t}|_{ng}$ to be a decreasing function along the null geodesic until $r_c$ and remain small thereafter. However, this does not occur for any solution that has $\Lambda< 0.82 (H_0^F)^2$. This suggests the system is fixed so that the redshift drift is zero along the null geodesic for the present observer but the state is not permanent and later (or earlier) observers would not have zero redshift drift along their null geodesics. This type of behavior have been found from the $\Lambda$CDM model: (after the $\Lambda$ has started to dominate the Friedmann equation) there exists always one coordinate distance where the redshift drift is zero along each observer's null geodesic, but the coordinate distance changes in time \cite{SundellVilja2014}.

Imposing the redshift drift to be approximatelly zero along the null geodesic of the present observer does not induce smoothed inhomogeneous $\Lambda$LTB models that are compatible with the SNIa data. We note that this scenario does not include the $\Lambda$CDM models as a special case, because here $\Lambda \lsim 1.1 (H_0^F)^2$, whereas the $\Lambda$CDM model has $\Lambda \sim 2.1 (H_0^F)^2$. Moreover, the LTB case, where $\Lambda=0$, is unsuitable due to the vast inhomogeneity of $e(r)$.

\begin{figure}
\centering
\includegraphics[scale=1]{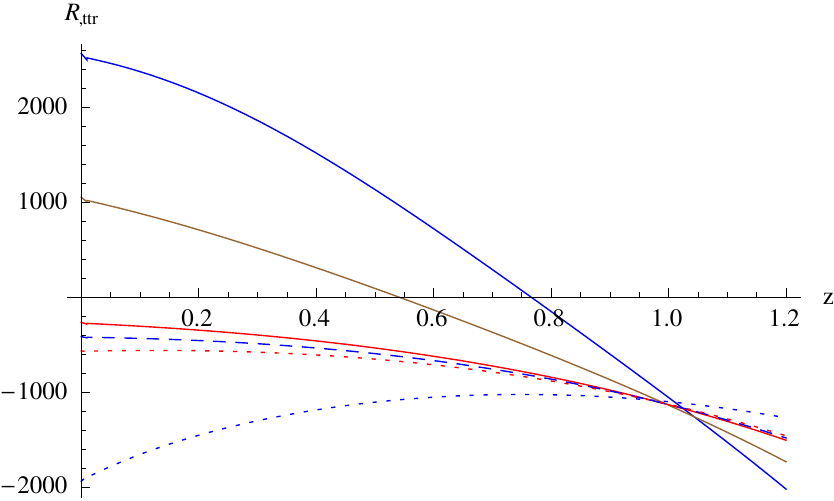}
\caption{\footnotesize  Solutions of the static universe scenario Eq.\ (\ref{bu}) with constraining Eq. (\ref{g}) corresponding to different $\Lambda$ values. In each solution, $H_0^F=70$, $\Omega_m^F=0.3$, $r_c=42$ Mpc, $\zeta=-492857$ and $M_0$ and $e_0$ are determined by the origin and AH conditions. }
\label{LTBk0}
\end{figure}

Let us consider another constraining equation, which imposes $R_{,t}|_{ng}\approx 0$ in the range $0.01\leq z\leq 1.2$:
\be{g}
R_{,t}|_{ng}=g(r),
\ee
where
\be{gdef}
g(r) =
  \left\{
  \begin{array}{ll}
 H_0^F -\gamma r&  r \leq r_c\\
 H_0^F -\gamma r_c \quad & r_c<r.
  \end{array}
\right.
\ee 
The parameter  $\gamma$ controls how close to zero $R_{,t}|_{ng}$ settles. We numerically explored the case $\Omega_m^F=0.3$ and $H_0^F=70$ by varying the parameters $M_0$, $\Lambda$, $e_0$, and $\gamma$. As earlier,  we used trial and error to find the parameter values which satisfy the origin and AH conditions. The results are not qualitatively dependent on how accurately $r_c$ is fixed to satisfy $z_c=0.01$ or how close to zero $R_{,t}|_{ng}$ settles: the solutions do not impose $dz/t_0\approx 0$, because the equality in Eq. (\ref{su}) is not met. This is illustrated in  Figure \ref{LTBk0}: if the equality in Eq. (\ref{su}) would hold, the curves $R_{,ttr}|_{ng}$  would set to constant for $z\in [0.01,1.2]$. We note that  this  scenario includes the LTB and the $\Lambda$CDM models as special cases.

In conclusion, we were unable to find the constraining equations which  satisfy the conditions for the static universe, given in Eq. (\ref{su}), while remaining compatible with the SNIa data. This is because the $\Lambda$LTB models do not include enough degrees of freedom; Eq. (\ref{su}) contains three conditions whereas the $\Lambda$LTB models left with one degree of freedom to constrain.

\subsection{The big universe}\label{biguniverse}

\begin{figure}
\centering
\includegraphics[scale=1]{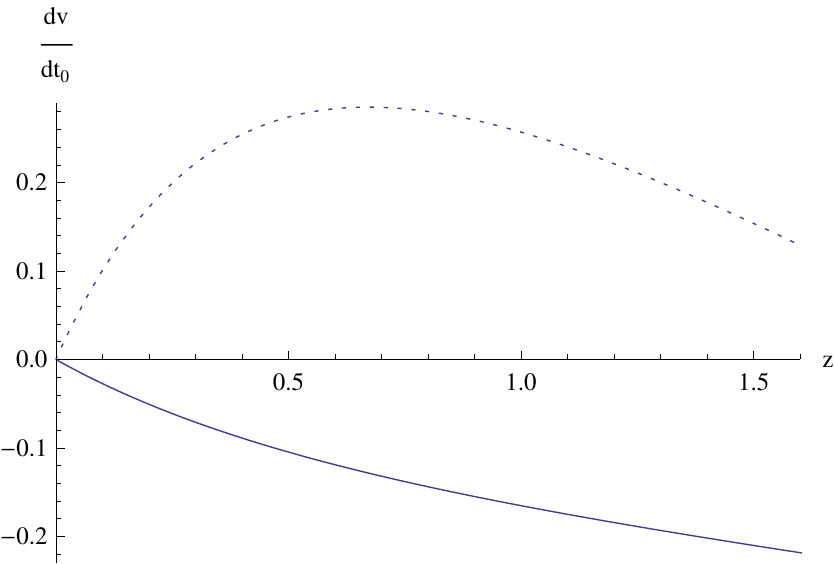} \,\,
\includegraphics[scale=1]{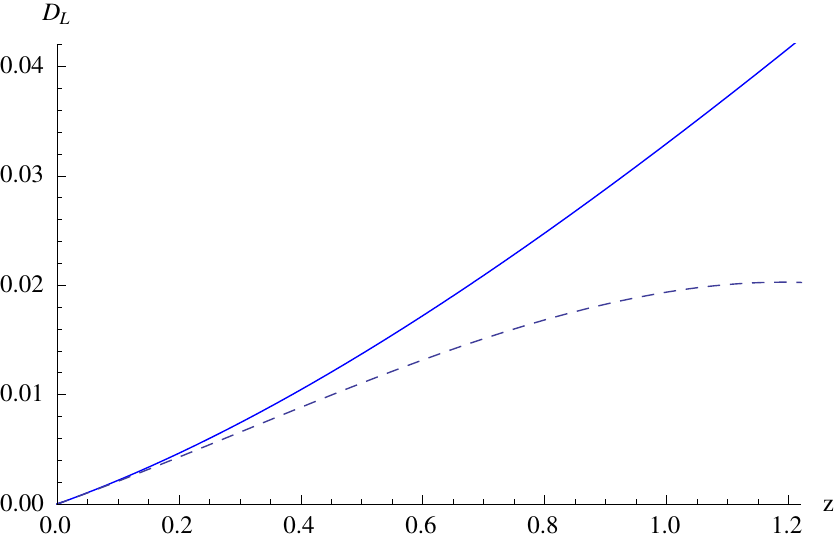} 
\caption{\footnotesize  The big universe scenario Eq.\ (\ref{bu}).  The left panel. The solid curve is the redshift drift with $e_2$  and parameters $\Omega_m^F=0.2$ and $H_0^F=50$  in units  cm s${}^{-1}$ yr${}^{-1}$. For comparison, the redshift drift of the $\Lambda$CDM model (characterized by parameters $\Omega_m^F=0.3$ and $H_0^F=70$)  is drawn in the figure (dotted curve).  The right panel. The solid curve is the luminosity distance  with $e_2$  and parameters $\Omega_m^F=0.2$ and $H_0^F=50$,    and the dashed curve represents how the  luminosity would appear in the case of homogeneous bang time.  }
\label{LTBk2}
\end{figure}

The big universe is defined in Eq. (\ref{bu}) and describes an ever expanding Universe without the cosmological constant, which can take place only if the universe is para- or hyperbolic (see Eqs.\ (\ref{R}), (\ref{R0}) and (\ref{R-})).
We studied the hyperbolic scenario numerically by solving the differential equations (\ref{rs}), (\ref{ng}), (\ref{gauge}), (\ref{DAr}) and (\ref{dRdr}) by imposing different functions for $e$, namely $e_i=e_i(r)$. According to our numerical investigations, the AH and the origin conditions fix all the parameters and functions for each set $(\Omega_m^F,H_0^F,e_i(r))$, i.e. there is only one solution for given $\Omega_m^F$, $H_0^F$ and $e_i(r)$ and the value of $M_0$ is obtained by trial and error. We explored the parameter space 
\be{region}
0.2 \leq \Omega_m^F \leq 0.4, \quad 50\leq H_0^F \leq 90
\ee
 using $e(r)=e_0$. The redshift drift decreases along with decreasing $\Omega_m^F$ and $H_0^F$, consequently we found the smallest redshift drift corresponding to $\Omega_m^F=0.2$ and $H_0^F=50$. The redshift drift along the line of sight was qualitatively independent on the parameters: $dz/dt_0$ was in each case monotonically decreasing with respect to $r$ as in Figure \ref{LTBk2}. It should be recognized that for each solution $t_b(r)$ was a decreasing function. This  is an adverse feature, because the regularity of the origin forbits changing the  age of the universe at the origin and thus the bang time function can be changed only outside the origin, thus smoothening the bang time function would make the universe younger. For example, if $t_b(r)$ is changed to constant, it is necessarily  $t_b(0)$.

\begin{figure}
\centering
\includegraphics[scale=1]{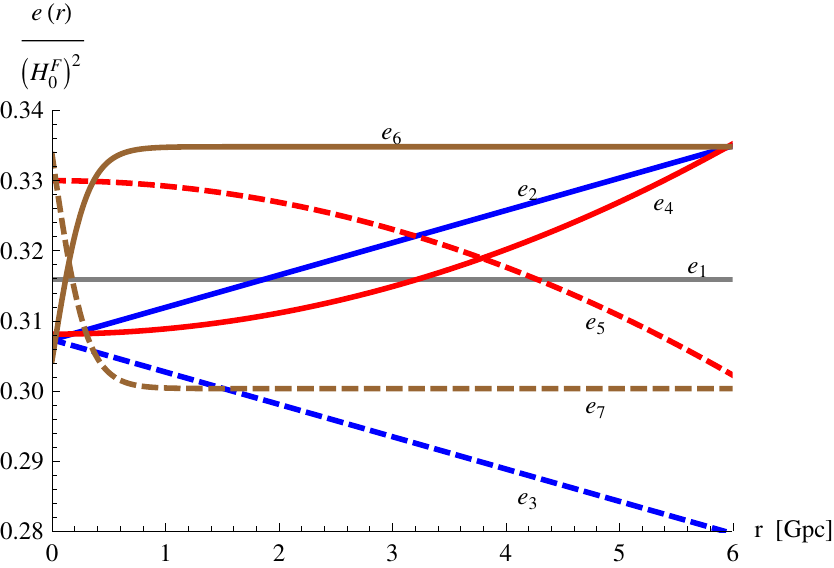} 
\caption{\footnotesize Representation of the $e(r)$ curves used to study the big universe (Eq.\ (\ref{bu})) characterized by parameters   $\Omega_m^F=0.2$ and $H_0^F=50$. The functions $e_i$ are explicitly presented in Table \ref{LTBt1}.}
\label{LTBk1}
\end{figure}

We investigated the case $\Omega_m^F=0.2$ and $H_0^F=50$ in more detail by imposing curves $e_i$ which are represented in Figure \ref{LTBk1} and in Table \ref{LTBt1}. The curves were constructed to respect the condition (\ref{econtrast}) so that $e_i|_{z=1.2}/e_0 \approx 0.9$ or $e_i|_{z=1.2}/e_0\approx 1.1$ for each $i \in \{2,7\}$ and $e_1=e_0$. The results are represented in Figure \ref{LTBk2} and in Table \ref{LTBt1}. The redshift drift is weakly dependent on the curves $e_i$, which is illustrated in the table by its values at $z=1.2$. The $dv/dt_0$ obtained using $e_2$ is depicted in the left panel  and two luminosity distances are drawn in the right panel of  Figure \ref{LTBk2}. The solid line corresponds to the luminosity distance where the functions $t_b=t_b(r)$, $t=t(r)$, $z=z(r)$, $e=e_2$, obtained by solving the Eqs. (\ref{rs}), (\ref{ng}), (\ref{gauge}), (\ref{DAr}), (\ref{dRdr}) and $e=e_2$,  whereas the dashed curve corresponds to the luminosity distance obtained otherwise the same way, but the bang time function is replaced a constant function. The two luminosity distances on Figure \ref{LTBk1} differ from each other considerably but the reason for this is clear: as discussed earlier, the two models have to have the same age in the origin and because  Eqs. (\ref{rs}), (\ref{ng}), (\ref{gauge}), (\ref{DAr}), (\ref{dRdr}) and $e=e_2$ yield a decreasing $t_b(r)$,  setting $t_b(r)$ to constant makes the universe younger. Furthermore, the set of equations yield a decreasing $t_b(r)$ for each $e_i$, $i \in \{1,7\}$.  Therefore, the system cannot be close to a state where the redshift drift is zero and we conclude this scenario is enable smoothening the inhomogeneity of this LTB model.

The imposed $\Lambda$CDM luminosity distance has the disadvantage that it dictates $e_0\sim 0.3$, whereas the corresponding values vary in the literature. In particular, larger $e_0$ values might reduce the redshift drift along the null geodesic for the following reason. The limit $R|_{ng} \to \infty$ in Eq. (\ref{bu}) could be replaced by inequalities $-M_{,r}(r)/R|_{ng}^2\ll 0.1$ cm s${}^{-1}$ yr${}^{-1}$ and $-2M(r) R_{,r}|_{ng}/R|_{ng}^3\ll 0.1$ cm s${}^{-1}$ yr${}^{-1}$. On the other hand, Eq. (\ref{R10}) implies that increasing $e_0$ decreases the ratio $M(r)/ R^3$ at the origin. Furthermore, this state should remain outside the origin if $e(r)$ is sufficiently homogeneous. To analyze this scenario further, we should omit Eq. (\ref{DAr}) and replace it with some other equation to close the system. However, this would require fitting SNIa data to ensure the compatibility of the models with the dimming of supernovae, which is out of the scope of this work. Therefore, this scenario remains without further analysis in this paper.

We studied the parabolic scenario analogously by imposing $e=0$. The origin conditions imply $M_0=(H_0^F)^2/2$ and according to our numerical evaluation  the AH condition is satisfied for all $\Omega_m^F \in [0.2,0.4]$ and $H_0^F \in [50,90]$, i.e. there is only one solution for given $\Omega_m^F$a and $H_0^F$  and the value of $M_0$ is obtained without trial and error. However, the redshift is not a monotonically increasing function with respect to $r$ and $z=1.2$ was not achieved for any $\Omega_m^F \in [0.2,0.4]$ and $H_0^F \in [50,90]$, which makes the scenario inviable.  

\subsection{The bizarre universe}

 The conditions characterising the bizarre universe are given in Eq.\ (\ref{biu}), stating that the redshift drift can converge to zero if $R$ and $1/R_{r}$ tend to infinity. However, we  show   this cannot take place at the AH and hence it is not credible to hold at  $z\in [0.01,1.2]$ either.

Taking the AH condition (\ref{AHc}) and substituting Eq.\ (\ref{gauge}) in, yields
\be{M0}
M_0=\frac{R}{2r^3}\left(1-\frac{\Lambda}{3}R^2 \right)
\ee
 Substituting this in the right hand side of Eq.\ (\ref{Rttr}), the  first term becomes
\be
-\frac{3M_0r^2}{R^2}=-\frac{3}{2Rr}+\frac{\Lambda R}{2 r},
\ee
which converges to infinity along with $R$, whereas the other terms in (\ref{Rttr}) tend to zero if $R\to \infty$. Thus, the redshift drift converges to infinity as $R\to \infty$ at the AH.

\section{Conclusions}\label{conclusions}

We have studied the possibility of constructing less inhomogeneous LTB models than usually presented in the literature to gain models viable to overcome the fine-tuning and observational issues of the LTB models. Moreover, we included non-zero cosmological constant in the equations to cover a more general class of models. We imposed the models to mimic the $\Lambda$CDM luminosity distance to ensure good fit to SNIa data. The new method constraining the degrees of freedom in these models is the novelty of our approach. In addition to constrain them with ansatzes, we considered also well-motivated constraining equations. The idea behind the constraint was that if the luminosity distance  does not change in time (the redshift drift is zero), we will not be able to know how long it has been so and therefore we do not know the age of the universe. Consequently, if we find one solution with a specific age profile, the same solution should apply also for less inhomogeneous and older age profiles. We found three scenarios viable to include smoothed inhomogeneity models, which we refer as the big universe, the static universe and the bizarre universe. We investigated these scenarios further using analytical and numerical methods.

The $\Lambda$LTB models have not been investigated without priors from the $\Lambda$CDM or LTB models. In Refs. \cite{Redlich2014,Valkenburg2014} the aim is to establish the Copernican principle, whereas in \cite{MarraPaakkonen2010} aimed to find the best fit $\Lambda$LTB model. However, they fix the steepness of the void to be related to the radius of the void so that a good fit between the SNIa data and the LTB models is guaranteed. This can bias the results to favor the LTB models (and $\Lambda$CDM models when the void is small enough) over the $\Lambda$CDM models with an unexpected value of $\Lambda$. Therefore, the results of our analysis of the $\Lambda$LTB cannot be anticipated, even though we use the $\Lambda$CDM luminosity distance which can be accounted as a $\Lambda$CDM prior.

The big universe scenario can take place only in the absence of the cosmological constant, hence it includes only the LTB models. There was no obvious way to find constraints that would lead to smoothed inhomogeneity. Thus, we studied how different luminosity distances and curvature profiles affect the smoothening. We found that they have very little effect and  we were unable to find less inhomogeneous models compared with those presented in the literature within this scenario. Two main reasons were identified: the redshift drift was not  zero and we were unable to modify the solutions so that they would become both less inhomogeneous and older at the same time. The latter reason arises because the age of the universe is fixed at the origin and   the bang time function was a decreasing function in  all of the implemented numerical solutions. Thus, we cannot reduce the inhomogeneity by making the universe older, but making it younger.  Imposing the universe to be younger means less time for the universe to evolve and is the opposite we were looking for to support our idea behind the scenario. However, we were unable to investigate the case of  the curvature dominated universe due to the limitations of our assumptions. Simple analytical considerations imply that increasing the curvature would decrease the redshift drift.  On the other hand, there is no guarantee that the redshift drift becomes small enough and furthermore, these models are required to have an increasing bang time function, unlike all the solutions we obtained for the dust dominated scenarios. Therefore, we reckon that smoothed inhomogeneity is hardly obtained with this scenario.

\begin{table} \caption{The results of the numerical analysis of the big universe model characterized by Eq.\ (\ref{bu}) and the parameters $\Omega_m^F=0.2$ and $H_0^F=50$. The AH is located at $z_{AH}=1.76307$, where $R_{AH}=2.67827$ Gpc and the AH condition is satisfied for each $e_i(r)$ by the accuracy $\sim 10^{-6}$. } \label{LTBt1} 
\begin{center}
  \begin{tabular}{| c | c | c | c | c | c | } 
    \hline 
 $e_i(r)$  &   $e_0/ (H_0^F)^2$  & $M_0/(H_0^F)^2$ & $r_{AH}\quad$ [Gpc] & $\frac{dz}{dt_0}|_{z=1.2} \quad $[km s${}^{-1}$Mpc${}^{-1}$] & $\frac{dv}{dt_0}|_{z=1.2} \quad $[cm  s${}^{-1}$ yr${}^{-1}$]    \\ \hline
 $e_1=e_0$ & 0.3159  &0.1841  & 6.39322&   $-13.3192$ &$ -0.185622$ \\ \hline
 $e_2=e_0(1+4.5 r)$ &0.3073  & 0.1927 &6.29624  &   $-13.2216$ &$-0.184261$ \\ \hline
$e_3=e_0(1-4.5 r)$  &0.3310  & 0.1690 & 6.57928 & $-13.4808$ &$-0.187874$\\    \hline
  $e_4=e_0(1+220 r^2)$ &0.3081  & 0.1919 & 6.30601   & $-13.2344$  &$-0.18444$\\ \hline
 $e_5=e_0(1-210 r^2)$  & 0.3300 & 0.1700 & 6.56542  & $-13.4373$ &$-0.187268$\\ \hline
$e_6=e_0(1+\tanh[1000 r]/10)$  &0.3044  &0.1956  &6.26518   & $-13.3443$ &$-0.185972$\\    \hline
$e_7=e_0(1-\tanh[1000 r]/10)$  & 0.3337  & 0.1663 &6.61468 &$-13.3524$ &$-0.186084$\\   \hline
  \end{tabular} 
\end{center} 
\end{table}

The static universe was of particular interest because we were able to provide novel type of constraining equations for the free functions of the $\Lambda$LTB models and this scenario included the LTB models and radially inhomogeneous $\Lambda$CDM models. We considered two constraining functions: one imposes the redshift drift to be zero along the null geodesic in the range $0.01\leq z\leq 1.2$ and the other sets the time derivative of the angular diameter distance close to zero in the same range. The numerical investigations implied that the scenario with the former constraining function is plagued by an enormous inhomogeneity of  curvature (the LTB model is included in these scenarios) or cannot become static ($R$ does not tend to a constant in time for any $r_c\leq r \leq r_{AH}$.  Furthermore, it turns out that the radially inhomogeneous $\Lambda$CDM model is excluded from these scenarios, because the maximum dark energy density at the symmetry center is $\approx 0.37$, whereas for the $\Lambda$CDM model it is  $\approx 0.7$. On the other hand, the latter constraining function was unable to make the redshift drift disappear, regardless of the value of $\Lambda$ (this scenario included both the LTB and the $\Lambda$CDM models as special cases). We were unable to impose both constraining equations simultaneously due to the lack of degrees of freedom in the $\Lambda$LRB models and conclude this to be the main reason why the static universe does not  include smoothed inhomogeneous models.

The bizarre universe has the peculiar feature that it is expanding in the angular direction while contracting in the radial direction. We analytically showed that this cannot hold at the apparent horizon at $z\approx 1.6$. Therefore, it appears implausible to hold at $z\in [0.01,1.2]$ either.

In summary, the inhomogeneity of the $\Lambda$LTB models compatible with the SNIa data cannot be substantially reduced by employing the method used here. Therefore, the observational and fine-tuning issues of the LTB models remain unsolved and we presume they cannot be solved using the $\Lambda$LTB models. However, the novel method introduced here can be applied to more general models, like the Szekeres models or the LTB models accompanied with a dynamical dark energy component. Moreover, more general models  can be expected to give positive results, albeit it is not possible to reduce the inhomogeneity in the $\Lambda$LTB models; more general models contain more degrees of freedom and at least by that part are better candidates than the  $\Lambda$LTB models. We shall leave the generalization to future work and expect our method  a powerful theoretical tool for probing models without data fitting.

\acknowledgments

We would like to thank Roberto A.  Sussman for useful discussions comments on the draft. This study is partially (P. S.) supported by the Vilho, Yrj\3 and Kalle V\2is\2l\2 Foundation.



\appendix

\section{Origin Conditions}\label{originconditions}

Break down of the numerical procedure was encountered  at the origin when solving the system of Eqs.  (\ref{rs}), (\ref{ng}), (\ref{gauge}),  (\ref{DAr}),  (\ref{dRdr})  and  an additional equation chosen according to the situation.  We used linear approximation in the vicinity of the origin to avoid the numerical difficulties. Next, we shall present the linear approximations of the functions $R$, $e$, $t$ and $z$  with respect to the origin, where the additional equation is Eq. (\ref{f}). 

In this appendix, all the functions are evaluated along the null geodesic and the prime $'$ denotes differentiation along the null geodesic with respect to the argument $r$.

We began by defining $R(t|_{ng},r)=A(r) r$ and rewriting  Eqs.  (\ref{rs}), (\ref{ng}), (\ref{gauge}),  (\ref{DAr}),  (\ref{dRdr}) as
\bea{A-2}
0&=&-z'(r)+\frac{a_1+a_2 r e'(r)}{a_3}, \\ \label{A-2b}
0&=&t'(r)+\frac{R_{,r}|_{ng}}{a_3}, \\ \label{A-2c}
0&=&a_5   r A'(r)+a_5 a_7+a_6 z'(r), \\ \label{A-2d}
0&=&-r A'(r)-a_7+a_8 r t'(r)+R_{,r}|_{ng}, \\ \label{A-2e}
0&=&a_9-R_{,r}|_{ng}.
\eea
where
\bea{A-1}
a_1&=&\frac{(z(r)+1) \left(6 A(r)^2 e(r)+9 M_0 A(r)+\Lambda  A(r)^3 R_{,r}|_{ng}-3 M_0
  R_{,r}|_{ng}\right)}{\sqrt{3} A(r)^2 \sqrt{\frac{6 M_0}{A(r)}+\Lambda 
   A(r)^2+6 e(r)}}, \\
a_2&=&\frac{\sqrt{3} (z(r)+1)}{\sqrt{\frac{6 M_0}{A(r)}+\Lambda  A(r)^2+6 e(r)}}, \\
a_3&=&\sqrt{2 r^2 e(r)+1},\\
a_5&=&H_0^F (z(r)+1), \\
a_6&=&r A(r) H_0^F-\frac{1}{\sqrt{(z(r)+1)^3 \Omega _m^F-\Omega _m^F+1}}, \\
a_7&=&A(r), \\
a_8&=&\sqrt{\frac{2 M_0}{A(r)}+\frac{1}{3} \Lambda  A(r)^2+2 e(r)}, \\
a_9&=&\frac{3 M_0/A(r)^2+f(r)  [z(r)+1] \sqrt{2 r^2 e(r)+1}}{2   M_0/A(r)^3+\Lambda/3}.
\eea
We eliminate $R(t|_{ng},r)$ from the Eqs.\ (\ref{A-2})-(\ref{A-2e}) and  some simple manipulation yields
\bea{Aer}
e'(r)&=&\frac{a_5 a_9 \left(a_8 r-a_3\right)-a_1 a_6}{a_2 a_6 r}, \\ \label{Azr}
z'(r)&=&\frac{a_5 a_9 \left(a_8 r-a_3\right)}{a_3 a_6}, \\ \label{AAr}
A'(r)&=&\frac{a_9-a_7}{r}-\frac{a_8 a_9}{a_3}, \\ \label{Atr}
t'(r)&=&-\frac{a_9}{a_3}.
\eea

By definition, $z(0)=0$. Remembering that $R|_{ng}=D_A$ and  using the L'Hopital's rule, we find
\be{A2}
\lim_{r\to 0}\frac{D_A}{r}=\lim_{r\to 0}\frac{z'(r)}{H_0^F}.
\ee
In general, $z'(r)|_{r=0}\neq 0$, hence $R|_{ng}\propto r $ at the origin. On the other hand, because we fixed the gauge by choosing (\ref{gauge}) and $R|_{ng}/r |_{r=0}=1$, this gives $A(0)=1$ and furthermore $z'(r)|_{r=0}=H_0^F$.

The first term on the right hand side of Eq. (\ref{AAr}) is not singular at the origin if $\lim_{r\to 0}a_9-a_7 \propto r$. This takes place only if 
\be{Af0}
f(0)=\frac{1}{3} \left(\Lambda -3 M_0\right).
\ee
The lowest order approximation of Eq.\ (\ref{A-2}) (or (\ref{rs}))  with respect to origin yields
\be{AH0f}
H_0^F=\sqrt{2M_0+2e_0+\frac{\Lambda}{3}},
\ee
where $e_0=e(0)$. We used this equation to give $e_0$.

The limit $r\to 0$ of both sides of Eqs. (\ref{Azr})-(\ref{Atr}) can now be taken  and solving the obtained equations yield
\bea{AA1}
A'(r)|_{r=0}&=&-\frac{9 M_0 H_0^F}{2 \left(6 M_0+\Lambda \right)} \\
e'(r)|_{r=0}&=&\frac{H_0^F \left(3 (H_0^{ F})^2 \left(6 M_0 \left(3 \Omega _m^F-1\right)+\Lambda  \left(3 \Omega_m^F+2\right)\right)-2 \left(\Lambda -3 M_0\right){}^2\right)}{12 \left(6  M_0+\Lambda \right)} \\
t'(r)|_{r=0}&=&-1
\eea

It is now straightforward to see that Eq. (\ref{AH0f}) the lowest order approximation of Eq.\ (\ref{R1}), too.  The age of the  present-local universe, $T_0=t_0-t_{b_0}$,  is obtained by integrating Eq.\ (\ref{R1}) as 
\be{Aage}
T_0=\int_0^1 \frac{dx}{\sqrt{\frac{M_0}{x}+2e_0+\frac{\Lambda}{3}x^2}}.
\ee
The results are invariant of the value chosen for $t_{b_0}$, but choosing this fixes $t_0$ via $T_0$, and vice versa.

The values of $R$, $e$, $t$, $z$ and their derivatives are now given at the origin with respect to parameters $M_0$ and $\Lambda$. The parameters are to be chosen so that the AH condition (\ref{AHc}) is satisfied.

Analogous methods were used to linearize the equations with respect to origin regardless of the constraining function. In particular, in Section \ref{biguniverse},  equation for $e$ was imposed instead of Eq.\ (\ref{f}) to close the system. Moreover, the solution of $R(t,r)$ is known (see Eqs.\ (\ref{R})-(\ref{R-})), hence we were able to express $R_{,r}|_{ng}$ with respect to $M$, $e$, $t$ and $t_b$. Consequently, we linearized
 also $t_b$ with respect to the origin in an analogous manner than above.

\end{document}